\documentclass[sigconf]{acmart}
\acmConference[ICSE 2024]{46th International Conference on Software Engineering}{April 2024}{Lisbon, Portugal}

\AtBeginDocument{%
  \providecommand\BibTeX{{%
    \normalfont B\kern-0.5em{\scshape i\kern-0.25em b}\kern-0.8em\TeX}}}

\copyrightyear{2024}
\acmYear{2024}
\setcopyright{acmlicensed}\acmConference[ICSE '24]{2024 IEEE/ACM 46th
International Conference on Software Engineering}{April 14--20, 2024}{Lisbon,
Portugal}
\acmBooktitle{2024 IEEE/ACM 46th International Conference on Software
Engineering (ICSE '24), April 14--20, 2024, Lisbon, Portugal}
\acmDOI{10.1145/3597503.3639082}
\acmISBN{979-8-4007-0217-4/24/04}

\usepackage{url}

\usepackage{amsmath,amssymb,amsfonts}
\usepackage{algorithmic}
\usepackage{graphicx}
\usepackage{enumitem}
\usepackage{booktabs}

\PassOptionsToPackage{hyphens}{url}
\usepackage{hyperref}
\hypersetup{colorlinks, citecolor=blue, filecolor=pink, linkcolor=red, urlcolor=magenta }
\usepackage[inkscapelatex=false]{svg}
\usepackage{textcomp}
\usepackage{xcolor,colortbl}
\definecolor{mygray}{gray}{0.95}
\definecolor{magicmint}{rgb}{0.85, 0.96, 0.92}
\usepackage{subfigure}
\usepackage{pifont}
\newcommand{\cmark}{\ding{51}}%
\newcommand{\xmark}{\ding{55}}
\newcommand{\tabincell}[2]{\begin{tabular}{@{}#1@{}}#2\end{tabular}}
\newcommand{\hytt}[1]{\texttt{\hyphenchar\font=\defaulthyphenchar #1}}
\def\BibTeX{{\rm B\kern-.05em{\sc i\kern-.025em b}\kern-.08em
    T\kern-.1667em\lower.7ex\hbox{E}\kern-.125emX}}

\usepackage{lipsum}
\usepackage[htt]{hyphenat}

\begin{document}

\title{An Empirical Study on Oculus Virtual Reality Applications: Security and Privacy Perspectives}

\author{Hanyang Guo}
\affiliation{%
  \institution{Hong Kong Baptist University}
  \city{Hong Kong}
  \country{China}
}
\affiliation{%
  \institution{Sun Yat-sen University}
  \city{Zhuhai}
  \country{China}
}

\email{guohy36@mail2.sysu.edu.cn}

\author{Hong-Ning Dai$^*$}\thanks{$*$ Corresponding author.}
\affiliation{%
  \institution{Hong Kong Baptist University}
  \city{Hong Kong}
  \country{China}
}
\email{hndai@ieee.org}

\author{Xiapu Luo}
\affiliation{%
  \institution{The Hong Kong Polytechnic University}
  \city{Hong Kong}
  \country{China}
}
\email{csxluo@comp.polyu.edu.hk}

\author{Zibin Zheng}
\affiliation{%
  \institution{Sun Yat-sen University}
  \city{Zhuhai}
  \country{China}
}
\email{zhzibin@mail.sysu.edu.cn}

\author{Gengyang Xu}
\affiliation{%
  \institution{Hong Kong Baptist University}
  \city{Hong Kong}
  \country{China}
}
\email{21253277@life.hkbu.edu.hk}

\author{Fengliang He}
\affiliation{%
  \institution{Hong Kong Baptist University}
  \city{Hong Kong}
  \country{China}
}
\email{csflhe@comp.hkbu.edu.hk}
\renewcommand{\shortauthors}{Guo and Dai, et al.}
\begin{abstract}
Although Virtual Reality (VR) has accelerated its prevalent adoption in emerging metaverse applications, it is not a fundamentally new technology. On one hand, most VR operating systems (OS) are based on off-the-shelf mobile OS (e.g., Android). As a result, VR apps also inherit privacy and security deficiencies from conventional mobile apps. On the other hand, in contrast to conventional mobile apps, VR apps can achieve immersive experience via diverse VR devices, such as head-mounted displays, body sensors, and controllers though achieving this requires the extensive collection of privacy-sensitive human biometrics (e.g., hand-tracking and face-tracking data). Moreover, VR apps have been typically implemented by 3D gaming engines (e.g., Unity), which also contain intrinsic security vulnerabilities. Inappropriate use of these technologies may incur privacy leaks and security vulnerabilities although these issues have not received significant attention compared to the proliferation of diverse VR apps. In this paper, we develop a security and privacy assessment tool, namely the VR-SP detector for VR apps. The VR-SP detector has integrated program static analysis tools and privacy-policy analysis methods. Using the VR-SP detector, we conduct a comprehensive empirical study on 500 popular VR apps. We obtain the original apps from the popular Oculus and SideQuest app stores and extract APK files via the Meta Oculus Quest 2 device. We evaluate security vulnerabilities and privacy data leaks of these VR apps by VR app analysis, taint analysis, and privacy-policy analysis. We find that a number of security vulnerabilities and privacy leaks widely exist in VR apps. Moreover, our results also reveal conflicting representations in the privacy policies of these apps and inconsistencies of the actual data collection with the privacy-policy statements of the apps. Based on these findings, we make suggestions for the future development of VR apps.
\end{abstract}



\maketitle

\section{Introduction} \label{intro}
Virtual Reality (VR)~\cite{wohlgenannt2020} has recently received a boosted development. Diverse VR devices and VR systems have been developed by Meta (previously Facebook), Apple, Microsoft, ByteDance, Sony, HTC, etc. As reported by Fortune~\cite{Fortune}, the global VR market size is projected to grow from \$25.11 billion in 2023 to \$165.91 billion by 2030. The proliferation of VR devices and VR systems has also greatly driven the development of the metaverse, which emphasizes users' immersive experience in virtual worlds~\cite{Stylianos2022} and the real-time interactions with 3D models in a VR environment. 

Despite its rapid development, VR is not a fundamentally novel technology. VR's conceptual prototypes were established several decades ago by implementing a computer simulation system to generating 3D objects in virtual worlds. The recent development of hardware and software has fastened the adoption of VR technology. With the proliferation of VR devices and metaverse systems, a large number of VR apps have been developed and released. Most of these VR apps are running on top of off-the-shelf mobile operating systems (OS), such as Android (as well as its variants), Sony Orbis OS (originated from FreeBSD 9), and Apple visionOS (based on iOS). As a result, VR apps share common features with conventional mobile apps and also inherit their intrinsic deficiencies. For example, many VR apps run on top of Android OS with underlying VR devices (e.g., Meta's Oculus Quest 2 \cite{Santoso2022} and ByteDance's Pico 4 \cite{Hui2022}). Consequently, VR apps developed based on these VR devices are packaged as Android PacKage (APK) files. After decompiling APK files, these VR apps also generate similar files to Android apps, such as \texttt{\small AndroidManifest.xml}, resource files, java files, and so on.

Despite the similarity to Android mobile apps, VR apps have unique characteristics different from conventional mobile apps. (i) VR apps include not only Personally identifiable information (PII) data like Android apps, but also VR-specific PII data (e.g., VR device ID, Controller ID, and sensor ID). (ii) Many VR apps have been developed on top of specific VR platforms (e.g., Oculus Quest SDK) and 3D game engines (e.g., Unity, Unreal Engine, etc.) for achieving an immersive user experience in the 3D environment. (iii) VR apps request not only identity information data like Android mobile apps, but also extensive access to human biometrics, such as iris or retina scans, fingerprints, hand-tracking as well as face-tracking data, and voice. The use of new data types requires additional permission authentication, which can place new requests on the specification of configuration files, e.g., including new VR-related permission flags in \texttt{\small AndroidManifest.xml} in addition to conventional Android permission flags. (iv) VR apps have more descriptions of access norms for human biometrics in their privacy policies, sharply different from conventional Android apps.

Sharing common features of Android apps while possessing different features, VR apps are exposed to not only known security and privacy vulnerabilities inherited from Android apps but also emerging security and privacy risks. Unfortunately, a comprehensive study on VR apps is largely missing in the literature compared to the proliferation of VR apps~\cite{Huang2023}. Ignorance of these emerging security risks will greatly dampen users' enthusiasm for using VR apps and purchasing VR devices as well as VR services~\cite{Mustafa2018,Martin2023}. For example,  as discovered in Bigscreen (i.e., a famous virtual social VR app), a vulnerability allows strangers to perform various intrusive activities without the user's consent, such as turning on the user's microphone and eavesdropping on private conversations~\cite{Casey2021}.

In order to address the above issues, we propose a \underline{VR} \underline{S}ecurity and \underline{P}rivacy assessment tool for VR apps, called \textit{VR-SP Detector} for detecting potential security vulnerabilities and privacy risks. To the best of our knowledge, this is the first study on evaluating security and privacy issues of VR apps from code analysis and privacy-policy analysis. Despite a recent study (\textsc{OVRseen}) on analyzing privacy policies of VR apps~\cite{Trimananda2022}, it mainly focuses on network traffic rather than the implemented codes of VR apps. Our VR-SP detector integrates both static code analysis tools and privacy-policy analysis techniques, thereby effectively revealing security and privacy risks. Based on the VR-SP detector, we conduct a comprehensive empirical study on 500 popular VR apps (i.e., more than 3 $\times$ of \textsc{OVRseen}) from the largest VR app store SideQuest. 

After extracting the original APK file for each app after installing it in Meta Oculus Quest 2, we then decompile the APK to get not only the configuration file but also the source codes as well as the intermediate representation (IR) files. We next conduct a comprehensive analysis of the decompiled codes of each app. In particular, we conduct an analysis on the configuration file~\texttt{\small AndroidManifest.xml} of each app to get app basic information and permission usage information. We adopt pattern recognition techniques to detect Android-related general security and privacy vulnerabilities from Java files and the corresponding IR (i.e., Smali) files. We also use a taint analysis framework to detect private data leaks. Since most VR apps utilize Unity to achieve immersive environment rendering, biometric data capture, and even In-App Purchasing (IAP) service \cite{Carlos2019, Zuo2022}, we hence leverage a static binary analysis framework to detect the flows, biometric data usage, and IAP data in Unity-developed VR apps. 
At last, we adopt a privacy policy analysis based on natural language processing (NLP) to detect inconsistencies in the privacy policies of VR apps. We have obtained many insightful findings, on which, we provide some suggestions on the development of VR apps. 
In summary, the main contributions of this work are summarized as follows.
\begin{itemize}[leftmargin=*, topsep=0pt]
\item We propose an automatic security and privacy evaluation tool for metaverse-related VR apps. This tool can detect not only general vulnerabilities but also sensitive (PII and biometric) data leaks.
\item Our tool integrates static analysis and privacy-policy analysis technologies with consideration of the unique characteristics of security and privacy issues of VR apps. Specially, we detect security threats of VR apps by pattern matching and taint analysis from the decompiled code and use an NLP-based method to analyze privacy policies.
\item We run our tool on 500 popular VR apps and find that more than 95.40\% of the apps exist no root detection (No RD) vulnerability and 37.20\% exist insecure random generator (IRG) vulnerability. Moreover, 44.40\% of the apps invoke functions using biometric data though there are no requests in the manifest file. Moreover, 55.20\% of apps have no privacy policy and 11.00\% of apps have contradictory statements in privacy policies.
\item Based on the findings, we provide some advice on VR app development. We make our tool available at \url{https://github.com/Henrykwokkk/Meta-detector}.
\end{itemize}

\section{Background}
\subsection{Taxonomy of VR apps} \label{taxonomy}
VR apps in the metaverse context have different characteristics from conventional VR apps. Conventional VR apps usually create virtual content for a single user whose activities typically occur alone, e.g., simulating a single-user adventure, playing a single-player game, and watching VR movies alone. Differently, VR apps in the metaverse context emphasize the interaction among multiple users. Moreover, this kind of VR app typically creates virtual elements from real-world elements such as buildings, objects, and characters. Further, they also greatly extend virtual spaces from computer games to education, socialization, and online business activities. Considering news reports and research papers \cite{Cheng2022, Park2022, Thien2023, Lee2022}, we mainly consider the following five types of VR apps:
\begin{itemize}[leftmargin=*, labelindent=0pt,noitemsep,topsep=0pt]
\item \textbf{Virtual society} \cite{Cheng2022, Park2022}. These VR apps provide users with virtual spaces to interact and socialize with others. Typical apps include Rec Room, VRChat, etc.
\item \textbf{Gaming} \cite{Thien2023, Park2022}. A large body of VR apps are themed with games, such as Pavlov VR, Echo VR, etc.
\item \textbf{Art and culture} \cite{Lee2022}. Many VR apps support virtual museums, exhibitions, concerts, and other cultural and artistic events, such as Forbidden City Journey, Gravity Sketch, etc.
\item \textbf{Education} \cite{Park2022}. As a growing trend in the metaverse, VR apps can provide virtual classrooms, learning resources, and online courses, such as EngageVR, VR Anatomy, etc.
\item \textbf{Business and finance} \cite{Ning2021}. Many VR apps provide virtual stores, trading platforms, and financial services, e.g., Decentraland.
\end{itemize}

\begin{figure*}
\centering
\includegraphics[width=0.92\textwidth]{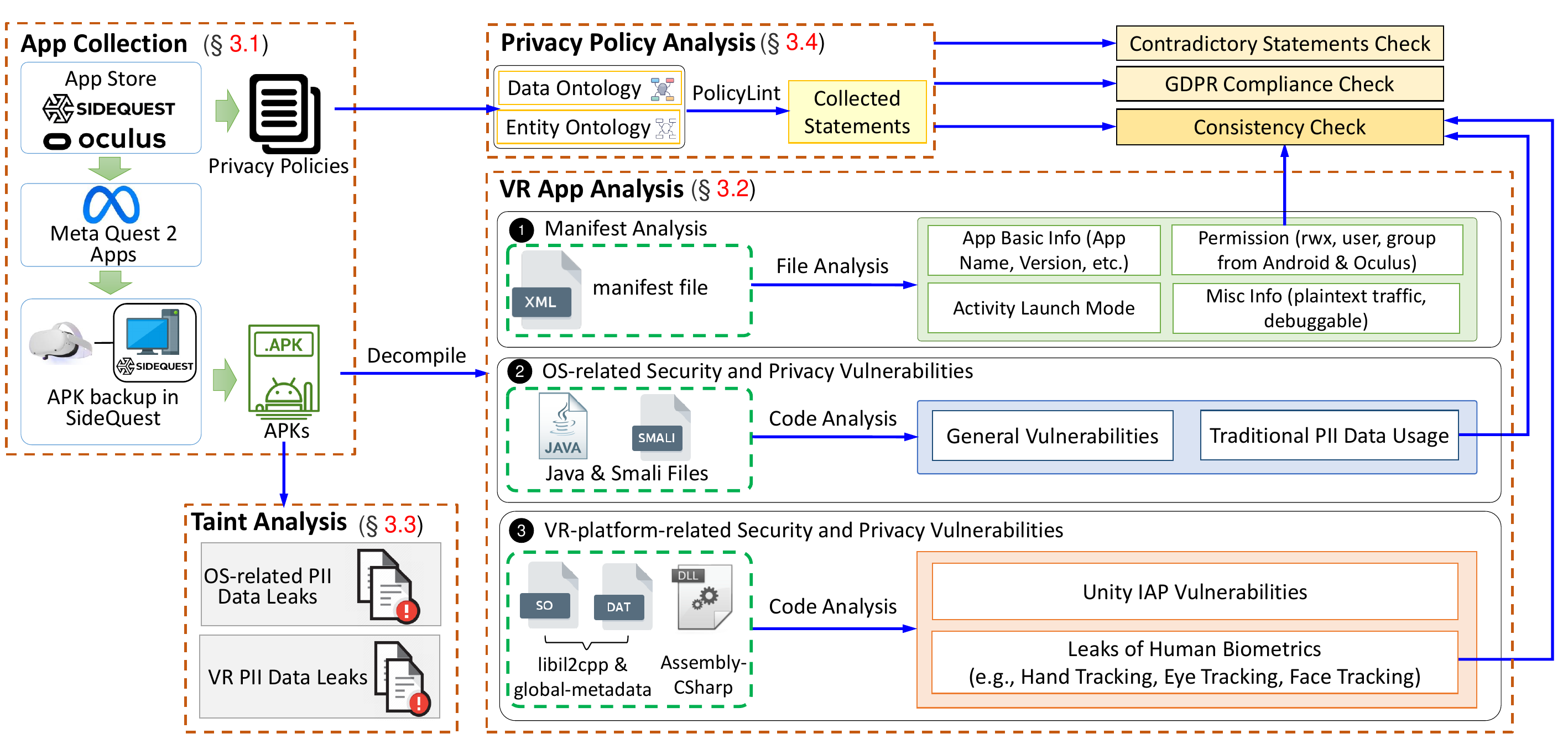}
\caption{The Overview of {\itshape VR-SP Detector}}
\label{overview}
\end{figure*}

\subsection{Security and Privacy Vulnerabilities of VR Apps}
\subsubsection{OS-related Security and Privacy Vulnerabilities} \label{OS}
Since a large body of VR devices run on top of Android OS or its variants, VR apps share some similar vulnerabilities with Android apps running on top of mobile devices like mobile phones although VR devices also have different features from mobile phones. We investigate the following security and privacy vulnerabilities, which are bestowed on new metaverse/VR features though they originated from Android security analysis \cite{Sun2021, Ma2019}.

\textbf{Insecure Flag Settings}: These vulnerabilities come from insecure flags in the configuration file \texttt{\small AndroidManifest.xml}, such as \texttt{\small allowBackup}, \texttt{\small debuggable}, and \texttt{\small clearTextTrafiic}.

\textbf{Dangerous Permission Usage}: These vulnerabilities are related to the misuse of dangerous Android permission requests, such as \texttt{\small location}, \texttt{\small camera}, \texttt{\small microphone}, etc. These vulnerabilities also exist in VR apps. Differently, we also consider permissions requests from other peripheral VR-related devices, e.g., controllers.

\textbf{PII Data Leaks}: Some of these vulnerabilities are related to PII data (e.g., user, name, password, email, and phone) leaks.

\textbf{General Vulnerabilities}: State-of-the-practice tools~\cite{LaMalva2020, Darvish2018, Shezan2017, Sun2021} also presented general vulnerabilities. We summarize them into the following nine categories in the VR context.

 (1) \textit{SQL Database Injection (SDI) in VR apps} \cite{Maha2022} means that an attacker can insert additional SQL statements to the end of a pre-defined query statement or input in an application to trick the database into executing an un-authorized query. 
 (2) \textit{Insecure certificate validation (ICV) in VR app client} \cite{Oltrogge2021} means that it might allow an attacker to spoof a trusted entity by interfering in the communication path between the host and client if a certificate is invalid or malicious.
 (3) \textit{Insecure random generator (IRG)} \cite{Hughes2022} means some insecure random number methods that can produce predictable values (e.g., virtual room passwords) as a source of randomness in a security-sensitive context. 
 (4) \textit{Insecure Webview Implementation (IWI)} \cite{Sai2019} means that the VR app allows loading HTML contents and HTML pages within the application. 
 (5) \textit{IP Disclosure (IPD)} \cite{Douglas2021} is a vulnerability that can be exploited by an attacker to obtain internal information from VR Apps' IP addresses. 
 (6) \textit{Remote Webview Debugging (RWD) of VR apps} \cite{Ang2021} means to enable webview debugging in VR apps. 
 (7) \textit{Unsafe sensitive data (such as user input by the virtual keyboard) cryptographic algorithms} include improper encrypt functions (IEF) \cite{Shao2014} and insecure hash functions (IHF)~\cite{Yoo2021}. 
 (8) \textit{Root Detection (RD) of VR devices} \cite{Zhang2015} means to detect the function usage that requires root access and check if the app asks to detect the rooted device. 
 (9) \textit{VR-related Tackers} \cite{Kollnig2021} include not only trackers in general Android apps (e.g., \textit{Google Firebase Analytics}) but also other VR-specified trackers (e.g., \textit{Unity3d Ads}).

\subsubsection{VR-platform-related Security and Privacy Vulnerabilities} \label{VR}
To achieve an immersive experience in the metaverse, VR apps typically implement diverse VR features, such as avatar modeling, 3D rendering, and 3D interaction. VR development frameworks, such as Unity \cite{unity} and Unreal Engine (UE) \cite{unreal} have been increasingly adopted. Both Unity and UE have been implemented in C++ though Unity has been partially implemented in C\#. Moreover, to capture users' location and movement, human biometrics, such as hand tracking, eye movement, face tracking, and body tracking have been collected and analyzed.  VR device manufacturers (e.g., Meta Oculus, HTC, Bytedance) provide developers with SDKs to achieve immersive VR features. However, the adoption of VR development frameworks and VR device SDKs inevitably causes new security vulnerabilities and privacy risks; this feature is \emph{sharply different from the development of conventional Android Apps}. In particular, we categorize security and privacy vulnerabilities related to VR development frameworks and VR device SDKs as VR-platform-related vulnerabilities, which are summarized as follows.
\begin{itemize}[leftmargin=*, labelindent=0pt,noitemsep,topsep=0pt]
    \item \textbf{New Permission Requests}: The use of new data (such as human biometrics) may introduce the problem of managing new permissions. Although device manufacturers provide new \texttt{\small <uses-permission>} tags for these new permission requests in the \texttt{\small AndroidManifest.xml} file, the technical regulation of using these permissions is still worth investigating.
    \item \textbf{Misuse of Human Biometrics}: Unity or UE-based frameworks include 3D rendering and exploit human biometrics by data collection APIs. For example, Oculus Unity SDKs provide some functions of collecting hand-tracking, eye-tracking, body-tracking, and face-tracking data\footnote{We mainly consider these functions: \hytt{\scriptsize OVRHand.OVRMesh.IOVRMeshDataProvider.GetMeshType}, \hytt{\scriptsize OVRBody.OVRSkeletonRenderer.IOVRSkeletonRendererDataProvider.GetSkeletonRendererData}, \hytt{\scriptsize VREyeGaze.CalculateEyeRotation}, and \hytt{\scriptsize VRFaceExpressions.ToArray}, etc.}. The misuse/abuse of such sensitive may cause severe security and privacy issues.
    \item \textbf{In-App Purchasing (IAP) Vulnerabilities}: With the prevalent adoption of IAP services in VR apps, it also incurs security risks. For example, players may purchase virtual items, such as virtual avatars, virtual currency, or NFT assets. Inappropriate authentication or authorization when purchasing virtual items may be the root cause of security vulnerabilities.
    \item \textbf{VR-specific PII Data Leak}: VR apps may suffer from the leakage risks of sensitive PII data, which include not only traditional PII data from Android systems but also VR-relevant PII data from VR devices (HMDs and peripheral devices). 
    \item \textbf{Privacy Policy Weakness}: Recently, several studies have been conducted to analyze the privacy policies of mobile apps (e.g., Android apps) to identify problems and verify their reliability~\cite{Luca2020, Jaime2020, Liao2020}, despite few studies on privacy policy analysis on VR apps. As newly emerging applications, VR apps are undergoing privacy policy weakness caused by both incomplete technical regulation originating from traditional Android development norms and new issues of VR devices in using and collecting users' privacy-sensitive data. 
\end{itemize}

\begin{table}[t]
\setlength{\abovecaptionskip}{0.1cm}
\caption{Keywords of Five VR App Categories}
\renewcommand{\arraystretch}{1.05}
\footnotesize
\begin{tabular}{ll}
\toprule
\textbf{Categories} & \textbf{Keywords} \\
\midrule
Virtual Society &  Social VR, Social Media, Virtual/Social Communications\\

Gaming & VR Games, VR Entertainment, Metaverse Games  \\

Art and Culture & Culture, Art, Museum \\

Education & Education, Teaching, Learning \\

Business and Finance & Finance, Business, Economic, NFT, DeFi \\ 
\bottomrule
\end{tabular}
\label{keyword}
\vspace{-0.4cm}
\end{table}

\section{Methodology of VR-SP Detector}
This section elaborates on the proposed VR-SP detector to analyze Oculus VR apps. Figure~\ref{overview} depicts the overall framework of the VR-SP detector. The proposed VR-SP detector works in the following steps: (1) App Collection, (2) VR App Analysis, (3) Taint Analysis, and (4) Privacy Policy Analysis, which are described as follows.

\subsection{App Collection} \label{data}
According to rankings and popularity, we collect the 500 most popular VR apps from both the official Oculus store and SideQuest. The latter is the most popular third-party store endorsed by Meta. Compared with the Oculus store, SideQuest receives a larger popularity and contains more apps. To obtain the original app files (not those from unauthorized parties), we download these apps from either the Oculus store or SideQuest. We then install them on Meta Quest 2 (aka Oculus Quest 2), one of the most popular VR devices. It is worth mentioning that our static analysis of these VR apps is not affected by VR devices even though some full-fledged features (e.g., eye-tracking and face-tracking) may require to be executed on higher-end VR devices (e.g., Oculus Meta Quest Pro). These apps are collected according to the categories specified in \S~\ref{taxonomy}. We obtained VR apps by the popularity and the keywords specified in Table~\ref{keyword}. We use the ``hot'' ranking in SideQuest as the popularity metric and consider only those free apps. At last, we collected 500 apps including 81 virtual social apps, 149 game apps, 115 art and culture apps, 115 education apps, and 40 business and finance apps (the distribution of those apps is given in \S~\ref{subsec:data-distr}).

After collecting apps, we then extract APK files from those apps. 
Since neither the Oculus official store nor SideQuest store provides a direct downloading link for APK files, we extract APK files by installing each app on the Oculus Quest 2 device. We first connect Oculus Quest 2 to the PC via a USB cable and install the app into Oculus Quest 2 via SideQuest. After that, we exploit the ``APK backup" function of SideQuest to extract the APK file corresponding to each app. For further privacy policy analysis (in \S~\ref{sec:privacy}), we also collect the policy statement of each of those 500 apps.

\subsection{VR App Analysis}\label{sec:app-analysis}

\begin{figure}[b]
\centering
\includegraphics[width=0.45\textwidth]{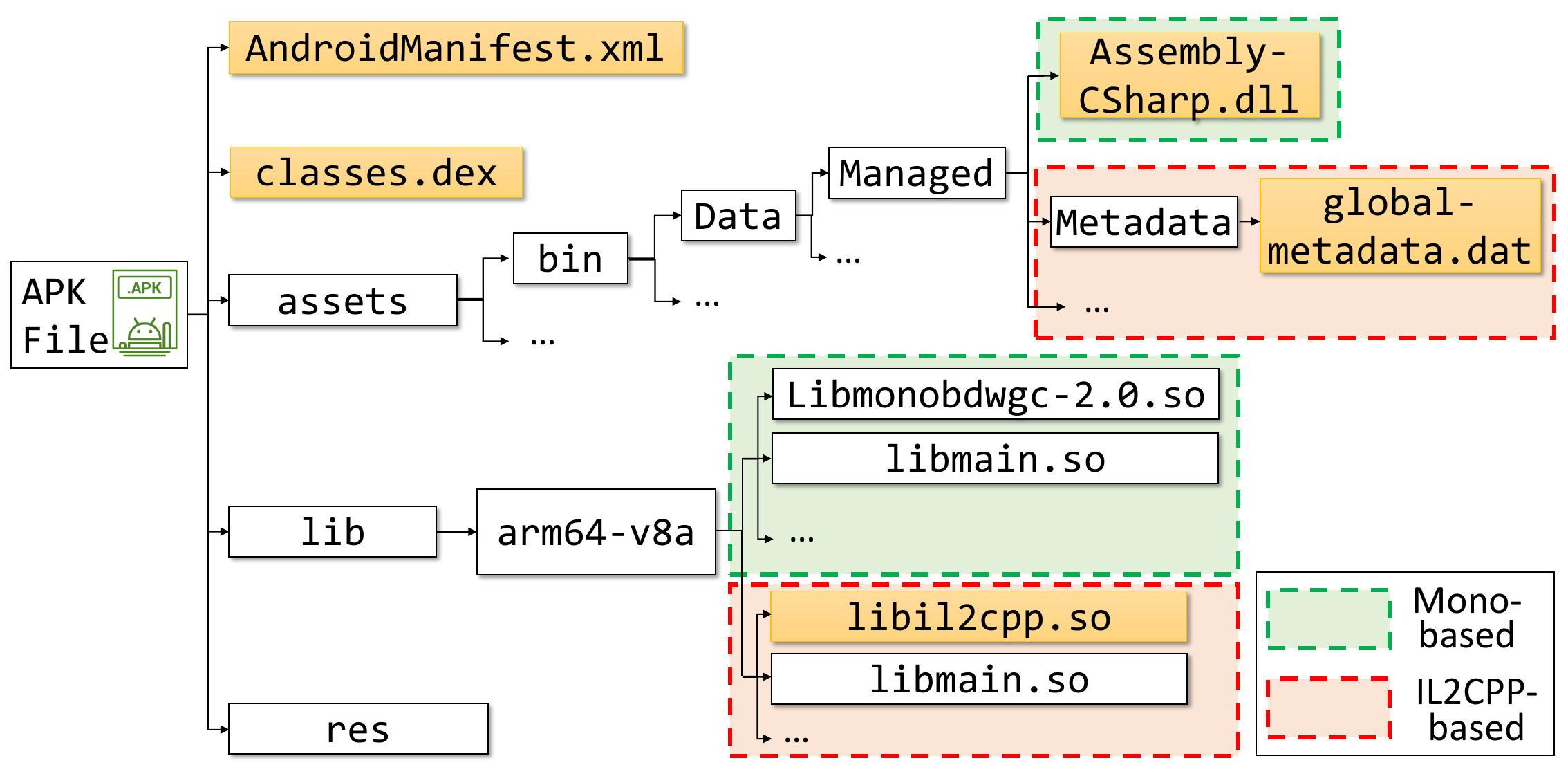}
\caption{APK Structure of VR App}
\label{APK Structure}
\end{figure}

We conduct VR app analysis based on the APK file extracted from the installed app. Since we cannot directly analyze a VR app, we first decompile its APK file. In order to attain a detailed program structure and development code information, we adopt Androguard~\cite{desnos2018},
 an open-source Python tool capable of extracting different kinds of information from the individual components of an APK file~\cite{Yang2023} to obtain configuration files, Dalvik bytecode, and source code. Figure~\ref{APK Structure} shows an example of the APK structure of VR apps.

From the decompiled files, we conduct \ding{182} Manifest Analysis, detect \ding{183} OS-related Security and Privacy Vulnerabilities, and identify \ding{184} VR-platform-related Security and Privacy Vulnerabilities on the decompiled code of each app as shown in Figure~\ref{overview}. The detailed analysis procedure is elaborated as follows.

\begin{table} [t] 
\caption{Pre-defined Rules of OS-related Vulnerabilities}
\begin{center}
\renewcommand{\arraystretch}{1.25}
\footnotesize

\begin{tabular}{cl}
\toprule
\textbf{Vulnerability Types} & \tabincell{c}{\textbf{Pre-defined Rules}} \\
\midrule
SDI &  \tabincell{l}{Search for dangerous SQL query method signatures \\in Smali files.} \\
\rowcolor{mygray}
ICV &  \tabincell{l}{Search for SSL or host verifier method signatures \\in Smali files.} \\
IRG &  \tabincell{l}{Search for random number generator method signatures \\in Smali files.} \\
\rowcolor{mygray}
IWI &  \tabincell{l}{Search for webview client and SSL errors received \\enable related method signatures in Smali files.} \\
RWD &  \tabincell{l}{Search for webview-debug-related method signatures \\in Smali files.} \\
\rowcolor{mygray}
IEF & \tabincell{l}{Search for cryptographic class and instantiation method\\ signature, and match insecure encryption keywords \\in Smali files.} \\
IHF & \tabincell{l}{Search for message digest class, instantiation method \\signature, and match weak hash pattern in Smali files.} \\
\rowcolor{mygray}
RD &  \tabincell{l}{Search for root and sudo access string.} \\
IPD & \tabincell{l}{Search IPV4, IPV6 and private address strings by \\regular expression matching.} \\
\rowcolor{mygray}
Trackers & \tabincell{l}{Find the tracker code signature that is included in the \\ tracker list in the Smali files.} \\
\hline
\end{tabular}
\label{Pre-defined Rules}
\end{center}
\end{table}

\subsubsection{Manifest Analysis} \label{Manifest Analysis}
After decompilation, we extract \texttt{\small AndroidMan}\-\texttt{\small ifest.xml} configuration file and parse it to extract key features of each VR app. In particular, we collect four types of features: 1) app basic information including \texttt{\small app\_name}, \texttt{\small package\_name}, \texttt{\small version\_code} and \texttt{\small sdk\_version} of the app; 2) permission information containing the permission requests, which include not only predefined permissions by Android but also those newly defined by VR devices (e.g., Oculus Quest 2); 3) activity launch mode, which refers to the start mode of activity in the task stack; and 4) miscellaneous information including \texttt{\small allow\_backup} and \texttt{\small use\_plaintext\_traffic} flag information; the settings of these flags have an impact on the security of data transmission of VR apps.

Regarding permission information, there are nine predefined permissions from Oculus (i.e., \texttt{\small com.oculus.permission}): \texttt{\small HAND\_TRACKING}, \texttt{\small RENDER\_MODEL}, \texttt{\small TRACKED\_KEYBOARD}, \texttt{\small USE\_ANCHOR\_API}, \texttt{\small FACE\_TRACKING}, \texttt{\small TOU\-CH\_CONTROLLER\_PRO}, \texttt{\small BODY\_TRACKING}, \texttt{\small EYE\_TRACKING}, and \texttt{\small DEVICE\_CONFIG\_PUSH\_TO\_CLIENT}. 
Moreover, 
the permission protection level can be divided into four categories: \textit{normal}, \textit{dangerous}, \textit{signature}, and \textit{signatureOrSystem} according to Android Development Documentation. We classify all the permissions defined by Oculus as \textit{dangerous} protection levels because they are all related to requests for sensitive information. With respect to activity launch mode, it can be used to detect the task-hijacking vulnerability in VR apps. Specifically, we identify the launch mode of activities that is \texttt{singleTask} without setting \texttt{taskAffinity} label, because this kind of activity may cause the task-hijacking vulnerability (i.e., StrandHogg)~\cite{Ren2015}.

\subsubsection{Detecting OS-related Security and Privacy Vulnerabilities} \label{OS-related}
We extract the decompiled partial Java files (without detailed variable names and method signatures) and Smali files that include Dalvik bytecode from \texttt{\small classes.dex} to detect some general security and privacy vulnerabilities indicated in \S~\ref{OS} in VR apps. Specifically, we use pre-defined patterns based on \cite{Sun2021} to detect potentially vulnerable methods (functions) and strings from Java codes or Dalvik bytecodes. We then search whether the function call paths exist. If the call paths or strings exist, we consider that the app being evaluated has this type of vulnerability. For example, suppose an app includes \textit{Cipher} and \textit{AES/ECB} keywords in the app. In that case, it may have IEF because the app uses the Electronic Code Book (ECB) mode in the cryptographic encryption algorithm. 
Since the same block of plaintext is encrypted into the same block of ciphertext in ECB, it may cause the leakage of encrypted messages~\cite{Elashry2009}. 
In summary, there are nine types of security and privacy vulnerabilities that we detect in VR apps. Table~\ref{Pre-defined Rules} summarizes pre-defined patterns. 
Moreover, we also extract the methods (functions) that collect and use PII data. We search method signatures by using PII keywords, such as \textit{user}, \textit{password}, \textit{username}, \textit{phone}, \textit{id}, and \textit{email} to identify PII data usage, which can be used for checking privacy policy consistency (to be depicted in \S~\ref{sec:privacy}).

\subsubsection{Identifying VR-platform-related Security and Privacy Vulnerabilities} \label{VR-platform-related}
This module focuses on detecting VR-platform-related security and privacy vulnerabilities, such as Unity IAP vulnerabilities and human biometrics leaks. As mentioned in \S~\ref{VR}, VR apps adopt game engines to achieve immersive features. We mainly consider Unity, which is the most dominating framework in VR software development~\cite{Nusrat2021}. We adopt a static native binary analysis tool~\cite{Zuo2022} to detect IAP vulnerabilities in VR apps developed based on Unity. There are two ways to run C\# programs on Unity. One is to compile the C\# code to .NET Common Intermediate Language (CIL) and use a Mono Virtual Machine (VM) to execute the CIL code at run time. This manner is called Mono-based (see green dash box in Figure~\ref{APK Structure}). The other is to further transfer CIL codes into C++ codes and then compile C++ codes into native binaries. This method is IL2CPP-based (see red dash box in Figure~\ref{APK Structure}). Corresponding to Mono-based and IL2CPP-based methods, APK files are named Mono-based and IL2CPP-based apps, respectively. Specifically, as for Mono-based apps, we extract the compiled logic code file \texttt{\small Assembly-CSharp.dll} and use a reversed tool called \textit{dnSpy} \cite{dnSpy}
to get the C\# source code. As for IL2CPP-based apps, we extract the compiled Unity binary file \texttt{\small libil2cpp.so} and the function mapping file \texttt{\small global-metadata.dat} from the decompiled APK file. Different from previous work~\cite{Zuo2022}, which only analyzed the IL2CPP-based apps, our proposed VR-SP detector focuses on both types of apps. Specifically, we use taint analysis to track the payment receipt data with the use of both the binary file and the mapping file. We define the method \texttt{\small UnityEngine.Purchasing.Product.get\_receipt} as the taint \textit{source} and the return value as tainted data. Regarding the detection of local-verification vulnerabilities (i.e., validating transactions only on the local server rather than asking the app store to verify the transaction), we define the method \texttt{\small CrossPlatformValidator.Validate} as \textit{sink}. If the tainted data is received while does not reach the network API, a \emph{local-verification vulnerability} is considered as detected. If the payment data is not sent externally (e.g., via a network API) and there is no local verification API involved, then a \emph{no-verification vulnerability} is considered as detected. 

Different from traditional Android apps, VR apps may collect human biometrics, such as eye location, hand coordinates, and so on. To tackle this emerging issue, we also trace the propagation of human biometrics in the proposed framework. Specifically, we extract the functions of collecting hand-tracking, eye-tracking, body-tracking, and face-tracking data described in \S~\ref{VR} according to the Oculus development document. We also conduct a taint analysis on those functions. Thereafter, we detect the usage of these biometric functions to check the risk of data leaks and also check whether their use is consistent with the permission request and the privacy policy statement (in \S~\ref{sec:privacy}).

\subsection{PII Data Leaks Identification}\label{sec:pii}
Since VR apps have high risks of PII data leaks, we conduct a taint analysis to detect sensitive information leaks based on data flow analysis. Firstly, we define VR-related PII data, which includes not only OS-related PII data (e.g., \textit{username}, \textit{phone}, \textit{email}, etc.), but also VR-specific PII data (e.g., \textit{VR device ID}, \textit{Controller ID}). Thereafter, referring to \cite{Steven2014}, we define the methods that transmit sensitive data as \textit{sources} (e.g., \texttt{\small getString(int)}, \texttt{\small getIntent()}, etc.). Moreover, we consider methods that may get sensitive data without user input, such as \texttt{\small getAddress()} for address information acquisition or \texttt{\small database.Cursor.getAllVisitedUrls()} for URLs queries in the database as \textit{sources}, too. The return values of the above methods are the tainted data \cite{Steven2014}. These tainted data may not be directly accessed with authentication but they may cause leaks if these data flow to a method can be accessed by unauthorized users. We define these modes as \textit{sinks}, such as \texttt{\small sendBroadcast()} and \texttt{\small sendDataMessage()}.

After defining the \textit{source}, \textit{sink}, and sensitive data, we search the lifecycle and method \texttt{\small callback} in the activity to construct the control flow graph (CFG). Based on CFG, we detect each defined source and taint the sensitive data. Then, we execute the data flow analysis to track the tainted data. If the tainted data flows from a source to a sink, we label it as a potential privacy-leak path. After conducting a backward ﬂow analysis, which aims to confirm the vulnerable code is reachable (i.e., not dead code) to reduce false positives, we identify that the app has a risk of privacy leak. For instance, if the return value of \texttt{\small getAddress()} flows to a \texttt{\small sink}, such as \texttt{\small sendDataMessage()} and any user or app that can access it without permission, we identify a privacy leak existing. 

\subsection{Privacy Policy Analysis} \label{sec:privacy}
The privacy policy is a complete and clear description of the practices of product and service providers in collecting, storing, using, and providing personal information to the public~\cite{Chang2019}. As mentioned in \S~\ref{VR}, few studies focus on privacy policy analysis on VR apps.
In our VR-SP detector, we implement a privacy policy analysis module to check the contradictory statements and the consistency between the privacy policy and the app analysis results. We use PolicyLint~\cite{Andow2019}, which transfers each statement in the privacy policy to plain texts and takes them as input. The output of the tools is collected statements formatted like \textit{\textless entity, action, data type\textgreater}, where \emph{entity} refers to the app or third-party platform that receives the privacy data, \emph{action} specifies the manner, in which entities process data and \emph{data type} is the type of privacy data. The categories of entity and data type are defined in~\cite{Trimananda2022}. For example, the privacy policy sentence ``\textit{We will collect your photo information and voice information for AI face pinching.}" can be input to PolicyLint to generate the collection statements \textit{\textless we, collect, voice information\textgreater} and \textit{\textless we, collect, photo information\textgreater}. We check privacy policies to see whether there are different collection operations for the same data type, i.e., both collection and non-collection. 

We also detect whether each privacy policy complies with General Data Protection Regulation (GDPR). GDPR was enacted by the European Union in 2018 and is one of the most well-known data privacy protection laws in the world \cite{Li2019}.  
We adopt a free online service called GDPRWise \cite{GDPRWise}
to conduct GDPR compliance checks.

\begin{figure}
\centering
\includegraphics[width=0.48\textwidth]{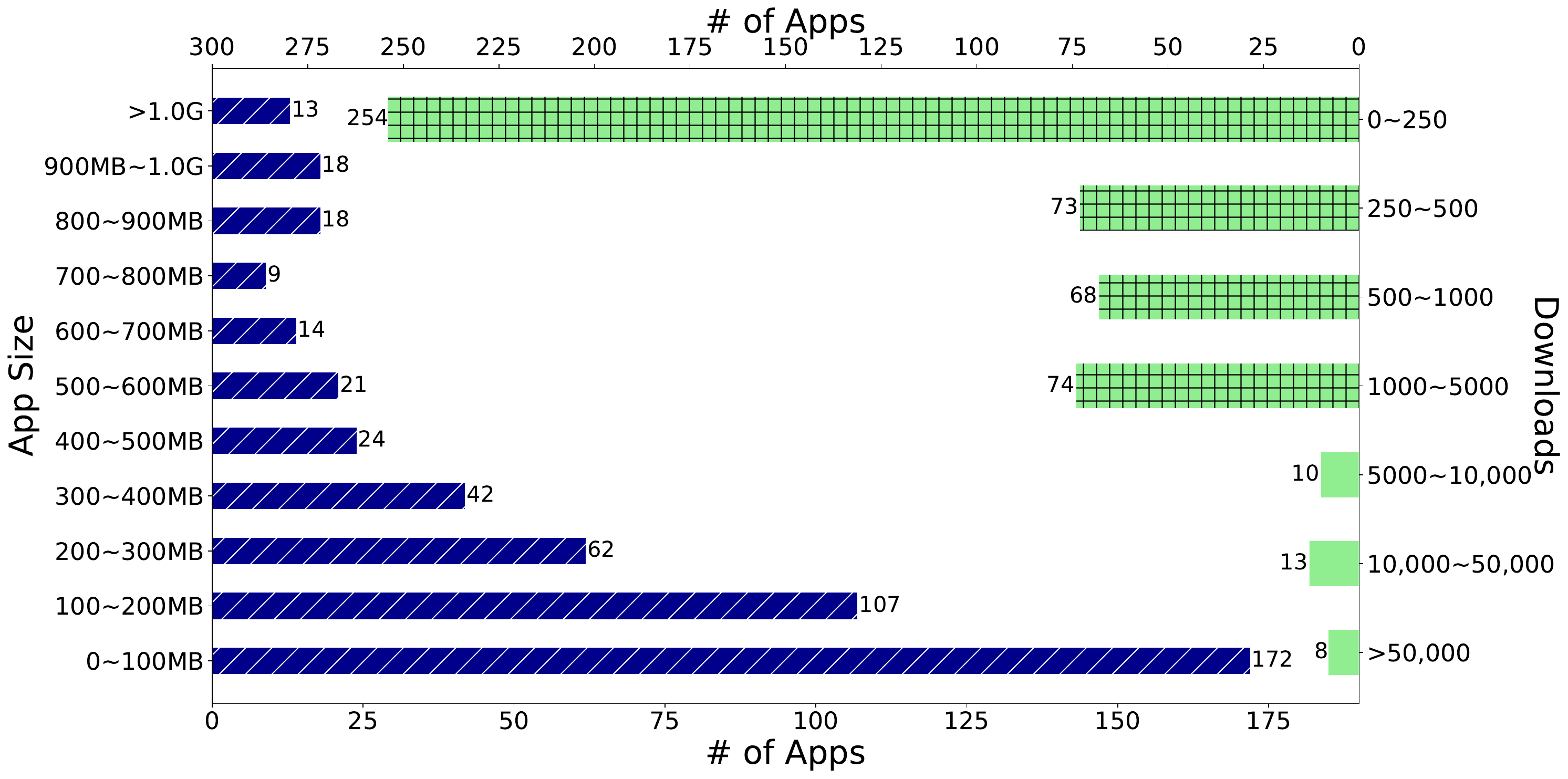}
\caption{App size \& Downloads Distribution}
\label{app-download-size}
\end{figure}

The consistency check module in our VR-SP detector includes three components. Firstly, we use the collected permission feature in the \textit{AndroidManifest.xml} file and adopt the permission request information in manifest analysis to make a consistency check for confirming the reliability of privacy policies. We extract the apps that request permissions \texttt{\small HAND\_TRACKING}, \texttt{\small FACE\_TRACKING}, \texttt{\small BODY\_TRACKING} and \texttt{\small EYE\_TRACKING}. We check whether their privacy policies have statements about the use of these sensitive data. We conduct a consistency check between the policy statements and the corresponding permission requests. Secondly, we use the result of decompiled Java and Smali code analysis. We extract methods that collect PII information by keyword searching and check whether their privacy policies have statements about the use of these PII data. Thirdly, we also conduct a consistency check between biometric collection function usage and privacy policy.

\section{Analysis and Results}\label{subsec:data-distr}
In this section, we evaluate the proposed VR-SP detector by conducting comprehensive experiments on the collected 500 VR apps. We mainly consider the following four research questions (RQs).
\begin{enumerate}[leftmargin=*,start=1,label={\bfseries RQ\arabic*:}]
\item What is the manifest vulnerability profile of VR apps?
\item What are VR apps' major OS-related security and privacy vulnerabilities?
\item What are VR apps' major VR-platform-related security and privacy vulnerabilities?
\item To what extent is PII data leaked?
\item How do the VR app developers comply with the privacy policies?
\end{enumerate}

\textbf{Data Preparation}. As mentioned in \S~\ref{data}, we collect 500 VR apps from five categories in the context of VR (the complete app list is given in our repository). Figure~\ref{app-download-size} summarizes the app size and download distribution of those 500 VR apps. It can be found that many apps are less than 100MB, which may own to limited space in current VR devices.
With respect to the number of downloads, most VR apps have been downloaded less than 5,000 times, implying that the development of VR apps is still in its early stage.


\begin{figure}[t]
\centering
\includegraphics[width=0.45\textwidth]{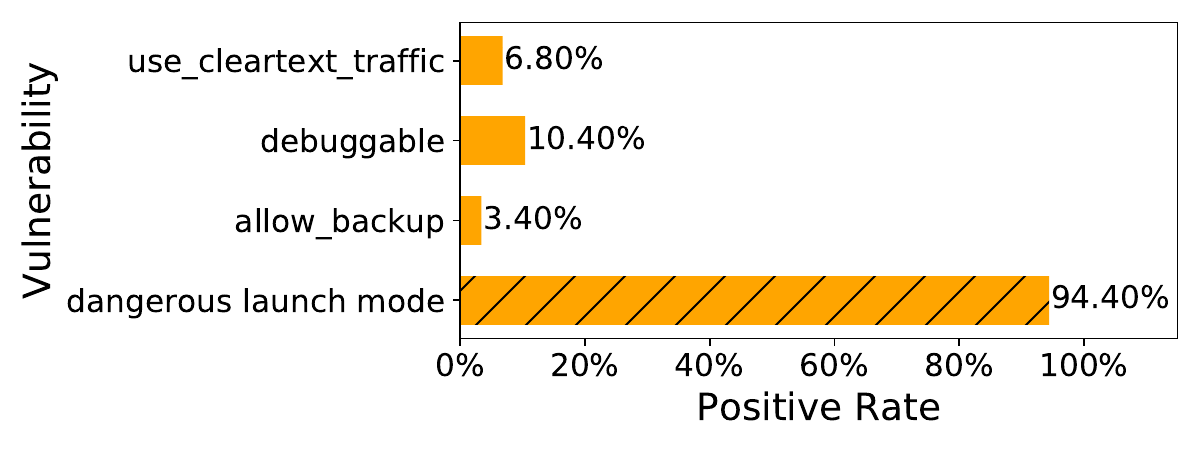}
\caption{Manifest Analysis Result}
\label{manifest}
\end{figure}

\subsection{RQ1: What is the manifest vulnerability profile of VR apps?}

Analyzing the manifest file of each app, we can find vulnerability risks, as shown in Figure~\ref{manifest}. It can be found that 94.40\% of the VR apps have activities having dangerous launch modes. For these activities, we further analyzed their specific contents and reported the results in Table~\ref{Non-standard Activities}. We find that most of the activities are \texttt{\small com.unity3d.player.\-UnityPlayerActivity} since these apps are developed based on Unity. Some other activities with dangerous launch modes are from third-party platforms (e.g., epicgame, google, etc.). Further, a root activity makes dangerous launch mode attributes be insecure since it is possible for other malware to read the contents of the calling intent. Table~\ref{Non-standard Activity app} shows examples of apps that have the most dangerous launch mode activities, where these apps are anonymized by MD5 encryption with the first five prefix letters. These apps need to be used with caution security risks.

\vspace{-0.2cm}
\begin{table}[h]
\caption{Dangerous Activities of VR apps}
\begin{center}
\small
\resizebox{0.85\linewidth}{!}{
\begin{tabular}{cc}
\toprule
\textbf{Types of Dangerous Activities } & \textbf{Amount} \\
\midrule
\texttt{com.unity3d.player.UnityPlayerActivity
} & 421 \\
\texttt{com.epicgames.ue4.GameActivity} & 32  \\
\texttt{com.google} Domain Activity & 12 \\
\texttt{com.epicgames.unreal} Domain Activity & 6 \\
\texttt{com.deepinc} Domain Activity & 2 \\
\texttt{com.microsoft} Domain Activity & 2 \\
\texttt{com.oculus} Domain Activity & 2 \\
Others & 13 \\
\bottomrule
\end{tabular}
}
\label{Non-standard Activities}
\end{center}
\end{table}

Compared with \textit{dangerous launch mode}, the percentage of apps containing other types of manifest vulnerabilities is relatively small. The positive rates of \textit{allow\_backup}, \textit{debuggable} and \textit{use\_cleartext\_tracffic} are 3.40\%, 10.40\% and 6.80\%, respectively. Enabling \textit{allow\_backup} and \textit{debuggable} causes a risk of coping and tampering with data from the device. This is even riskier in VR devices with human biometrics collected. Further, \textit{use\_cleartext\_tracffic} may cause a Man-in-the-Middle (MITM) \cite{Mauro2016} attack.

\begin{table}[h]
\caption{App Examples with Most Dangerous Launch Mode Activities}
\resizebox{\linewidth}{!}{
\renewcommand\arraystretch{1.15}
\begin{tabular}{cc}
\toprule
\textbf{App MD5 Prefix} & \textbf{Activity} \\
\midrule
\rowcolor{mygray}
\texttt{203fc}\textasciitilde 
& \tabincell{c}{\texttt{com.google.firebase.auth.internal.GenericIdpActivity} \\ \texttt{com.google.firebase.auth.internal.RecaptchaActivity} \\ \texttt{com.unity3d.player.UnityPlayerActivity}} \\
\texttt{d3458}\textasciitilde 
& \tabincell{c}{\texttt{com.google.firebase.auth.internal.GenericIdpActivity} \\ \texttt{com.google.firebase.auth.internal.RecaptchaActivity} \\ \texttt{com.unity3d.player.UnityPlayerActivity}} \\
\rowcolor{mygray}
\texttt{a0165}\textasciitilde 
& \tabincell{c}{\texttt{com.deepinc.liquidcinemasdk.SettingsActivity} \\ \texttt{com.deepinc.liquidcinemasdk.VideoSixGridActivity}}\\

\texttt{6e800}\textasciitilde 
& \tabincell{c}{\texttt{com.google.firebase.auth.internal.GenericIdpActivity} \\ \texttt{com.google.firebase.auth.internal.RecaptchaActivity} \\ }\\
\rowcolor{mygray}
\texttt{0163c}\textasciitilde 
& \tabincell{c}{\texttt{com.google.android.play.core.missingsplits.PlayCoreMissingSplitsActivity} \\ \texttt{com.unity3d.player.UnityPlayerActivity}}  \\

\texttt{4e336}\textasciitilde 
& \tabincell{c}{\texttt{com.pico.loginpaysdk.auth.TransferStationActivity} \\ \texttt{com.stormx.forbiddencityjourney.MainActivity}} \\
\rowcolor{mygray}
\texttt{9bbd5}\textasciitilde 
& \tabincell{c}{\texttt{com.pico.loginpaysdk.auth.TransferStationActivity} \\ \texttt{com.unity3d.player.UnityPlayerActivity}} \\

\texttt{a2114}\textasciitilde 
& \tabincell{c}{\texttt{com.epicgames.ue4.GameActivity} \\ \texttt{com.google.ar.core.InstallActivity}} \\
\bottomrule
\end{tabular}
}
\label{Non-standard Activity app}
\end{table}

\begin{figure}
\centering
\includegraphics[width=0.45\textwidth]{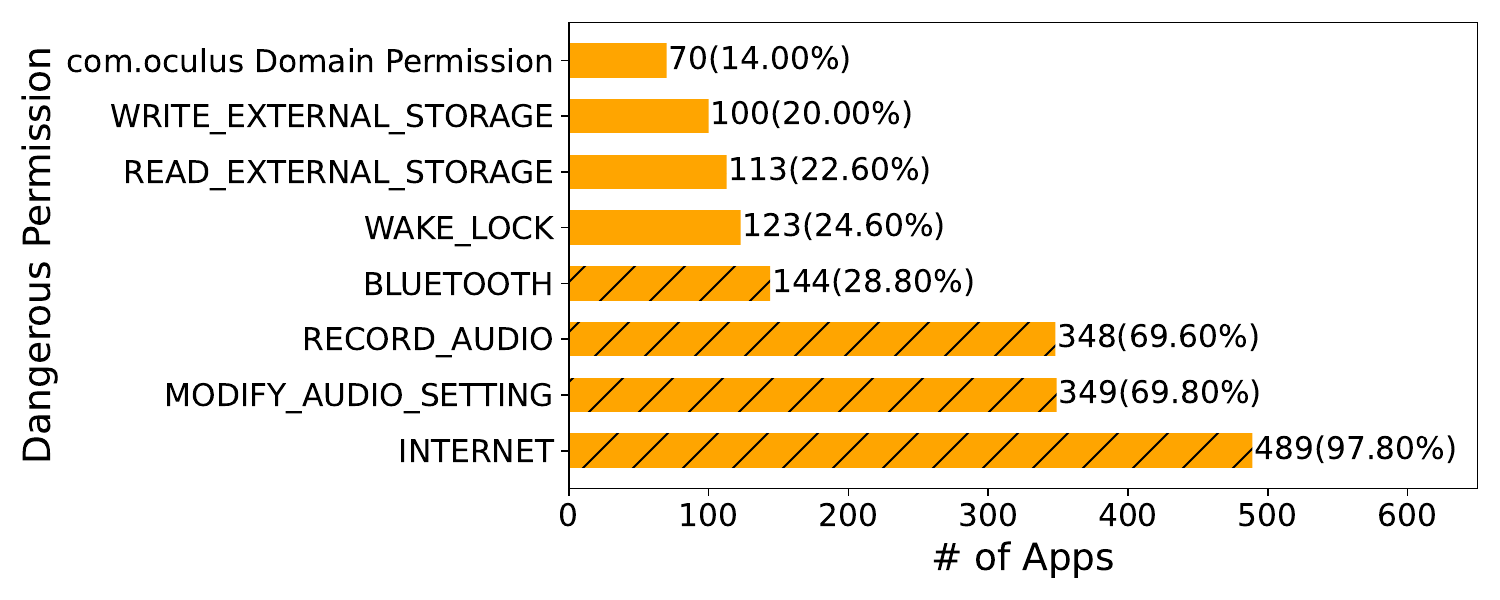}
\caption{Dangerous Permission}
\label{permission}
\end{figure}

We also count the most used dangerous permissions and the number of the used Oculus permissions mentioned in \S~\ref{sec:app-analysis}. As reported in Figure~\ref{permission}, most VR apps use INTERNET permission though some of them use permissions related to sound recording. This is because the social properties and immersive nature of VR apps require the Internet access and record the user's voice. Meanwhile, 70 apps use Oculus permissions to attain sensitive data. 
It is necessary to check whether dangerous permissions are secure before installing them. It is also important to manage app permissions by checking which permissions are allowed and declining if necessary.

\begin{center}
\fcolorbox{black}{gray!10}{\parbox{\linewidth}{\textbf{Answer to RQ1}: Most of the VR apps have a dangerous launch mode and sound-recording. Some apps have backup, debug, and network traffic misuse.}}
\end{center}

\begin{figure}[b]
\centering
\includegraphics[width=0.45\textwidth]{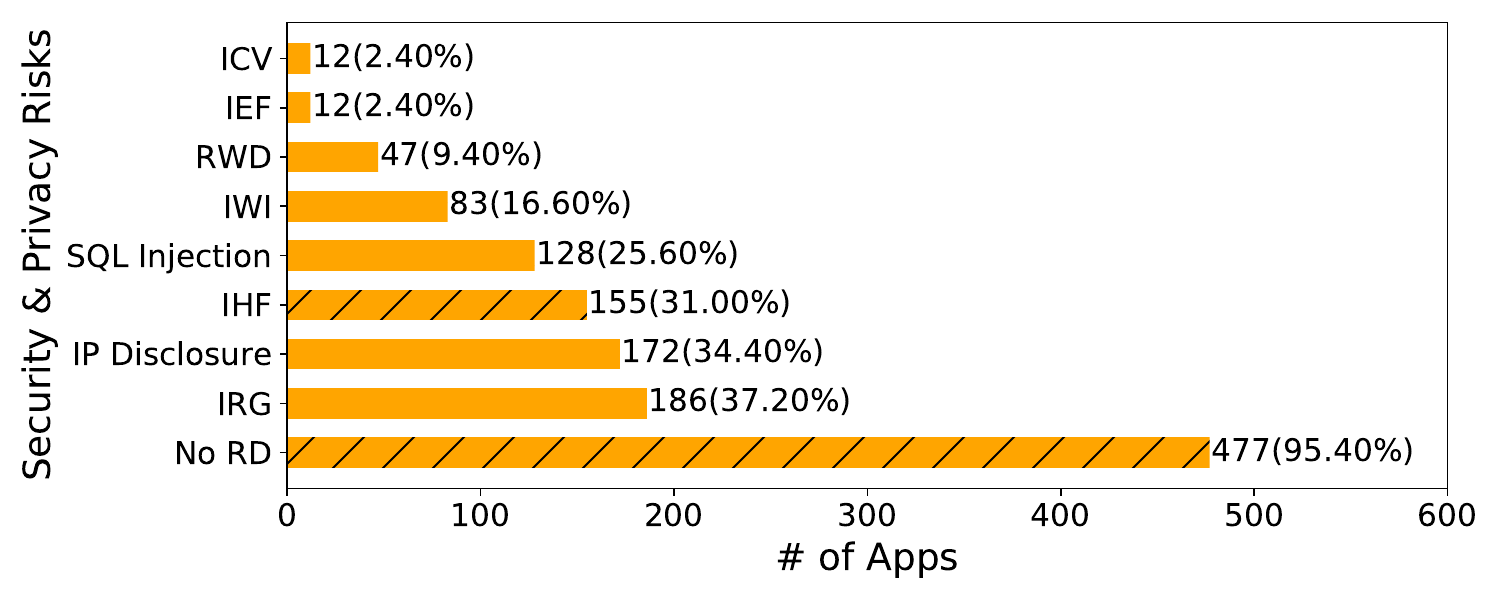}
\caption{Results of Code Analysis}
\label{codeanalysis}
\end{figure}

\subsection{RQ2: What are VR apps' major OS-related security and privacy vulnerabilities?}

As mentioned in \S~\ref{sec:app-analysis}, we detect security and privacy vulnerabilities from decompiled Java and Smali files. The results are shown in Figure~\ref{codeanalysis}. It can be found that most VR apps have the no-root-detection vulnerability of VR devices (477 apps), thereby leading to private information leakage caused by accessing databases with root privileges. Meanwhile, 186 apps have insecure random generators, which may cause private information (e.g., location) to be tracked by speculating random numbers on critical functions such as \texttt{\small generateDefaultSessionId} \texttt{\small ChangeWatermarkPosition} used in virtual social and business apps. In addition, IHF can also cause a serious security issue in VR apps. Attackers can exploit IHFs to conjecture users' input via analyzing typing activities on virtual keyboard~\cite{10179301}. 
Table~\ref{Security and Privacy Risks App} shows examples of apps (anonymized by MD5) having the most security and privacy risks. 

\begin{table}[t]
\caption{App Examples with Security and Privacy Risks}
\begin{center}
\resizebox{0.99\linewidth}{!}{
\begin{tabular}{ccccccccc}
\toprule
\textbf{App MD5 Prefix} & \texttt{cda01}\textasciitilde & \texttt{4dbd4}\textasciitilde & \texttt{0163c}\textasciitilde & \texttt{9bbd5}\textasciitilde & \texttt{58dea}\textasciitilde & \texttt{7d3df}\textasciitilde & \texttt{3aa85}\textasciitilde & \texttt{54208}\textasciitilde \\
\midrule
\textbf{SDI} & \cmark & \cmark & \cmark & \cmark & \cmark & \cmark & \cmark & \cmark\\
\textbf{ICV} & \xmark & \xmark & \xmark & \cmark & \xmark & \xmark & \xmark & \xmark\\
\textbf{IRG} & \cmark & \cmark & \cmark & \cmark & \cmark & \cmark & \cmark & \cmark \\
\textbf{IWI} & \cmark & \cmark & \cmark & \cmark & \cmark & \cmark & \cmark & \cmark \\
\textbf{IPD} & \cmark & \cmark & \cmark & \cmark & \cmark & \cmark & \cmark & \cmark \\
\textbf{RWD} & \xmark & \cmark & \cmark & \xmark & \xmark & \cmark & \cmark & \cmark \\
\textbf{IEF} & \cmark & \xmark & \xmark & \cmark & \cmark & \xmark & \xmark & \xmark \\
\textbf{IHF} & \cmark & \cmark & \cmark & \cmark & \cmark & \cmark & \cmark & \cmark \\
\textbf{No RD} & \cmark & \cmark & \cmark & \cmark & \cmark & \cmark & \cmark & \cmark \\
\bottomrule
\end{tabular}}
\label{Security and Privacy Risks App}
\end{center}
\end{table}

We also identify trackers used in VR apps by code analysis. Table~\ref{tracker} summarizes the most typical 15 trackers of VR apps. We observe that 108 VR apps use trackers, implying the prevalent usage of trackers in VR apps. Although the use of trackers can help VR app developers to provide users with customized services, it may expose users to the risk of privacy breaches. For example, \cite{Andow2020} indicated that sharing sensitive data with distinct advertisers (trackers), such as \textit{Unity3d Ads} is egregious.

\begin{table}[h]
\caption{Tracker Analysis}
\resizebox{0.99\linewidth}{!}{
\renewcommand{\arraystretch}{0.95}
\begin{tabular}{lc}
\toprule
\textbf{Trackers} & \textbf{\# of Apps} \\
\midrule
Google Play Billing Library / Service & 62 \\
Unity3d Ads & 40  \\
Google Firebase Analytics & 15 \\
GameAnalytics & 8 \\
Umeng Analytics & 3 \\
Umeng Common SDK logging & 3 \\
Bugsnag & 3 \\
Amplitude & 2 \\
Microsoft Visual Studio App Center Analytics & 2 \\
Microsoft Visual Studio App Center Crashes & 2 \\
Singular & 1 \\
AppMetrica & 1 \\
Branch & 1 \\
Bugfender & 1 \\
Google AdMob & 1 \\
\bottomrule
\end{tabular}
}
\label{tracker}
\end{table}

\begin{center}
\fcolorbox{black}{gray!10}{\parbox{\linewidth}{\textbf{Answer to RQ2}: Most VR apps have the no-root-detection vulnerability of VR devices. A significant number of apps have used trackers.}}
\end{center}

\subsection{RQ3: What are VR apps' major VR-platform security and privacy vulnerabilities?}

We next detect Unity IAP vulnerabilities and the usage of human biometrics. Regarding CIL to C++ Unity-based apps, we further obtain 332 apps from 500 apps. Depending on different approaches (Mono-based and IL2CPP-based), we obtain 124 Mono-based apps. We taint the Unity IAP function and biometric function. As for the results of Unity-based code analysis (i.e., C\# code analysis), we find that there are 28 apps adopting the Unity IAP function. According to the taint analysis, there exist IAP no-verification vulnerabilities in these 28 apps. Moreover, Table~\ref{biometric} shows that there are 224 apps (including Mono-based and IL2CPP-based apps) adopting biometric data collection functions while having no permission requests in the manifest file. Specifically, 209 apps use the hand-tracking function while indicating no request for that permission in the \texttt{\small Androidmanifest.xml} file. Among them, 62 apps exploit face-tracking functions, 59 apps invoke eye-tracking functions, and 63 apps call body-tracking functions. This implies that \emph{a significant number of apps do not adhere to the specifications of the Oculus VR app development documentation}. Biometric data collection can be enabled without user permission\footnote{This issue was raised in the Meta community forum, but no explicit answer has been given until July 2023: \scriptsize{\url{https://communityforums.atmeta.com/t5/Oculus-Quest-2-and-Quest/Unity-Oculus-Integration-bug-Hand-tracking-always-enabled-no/td-p/753132}}}. It exposes the risk of unknowingly stealing biometric data from users. It also indicates that Unity development framework for VR apps and the Oculus SDK have some potential vulnerabilities.

\begin{table}[t]
\caption{Inconsistency of Biometric Function Usage}
\resizebox{0.7\linewidth}{!}{
\begin{tabular}{cc}
\toprule
\textbf{Biometric Function} & \textbf{\# of Apps} \\
\midrule
Hand-Tracking Function & 209 \\
Eye-Tracking Function & 59  \\
Body-Tracking Function & 63 \\
Face-Tracking Function & 62 \\
\bottomrule
\end{tabular}
\label{biometric}
}
\end{table}

\begin{center}
\fcolorbox{black}{gray!10}{\parbox{\linewidth}{\textbf{Answer to RQ3}: Although only 5.60\% of VR apps have used Unity IAP functions, all of them have IAP no-verification vulnerabilities. 49.12\% of Unity VR apps have inconsistency between permission requests and biometric function usage, thereby causing leakage risks of human biometrics.}}
\end{center}

\subsection{RQ4: To what extent is PII data leaked?}
\begin{figure}[t]
\centering
\includegraphics[width=0.48\textwidth]{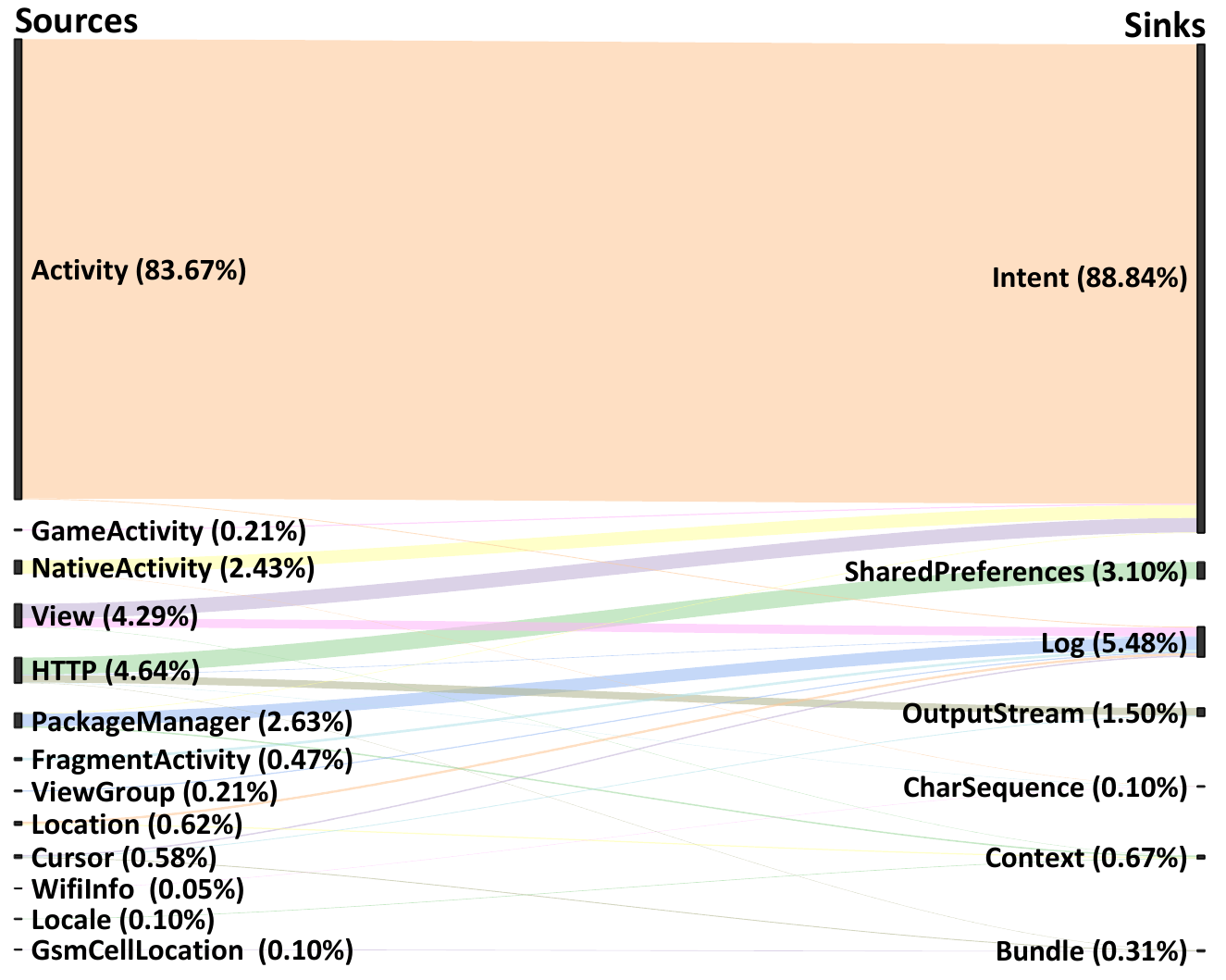}
\caption{Data Leaks Detection by Taint Analysis}
\label{flowdroid}
\end{figure}

We also adopt a taint analysis of PII data leaks. Figure~\ref{flowdroid} reports the results of the taint analysis for PII data leakage by calculating a percentage of the number of source-to-sink paths found in each VR app. It can be found that calling from Activity-related methods such as \texttt{\small Activity}, \texttt{\small NativeActivity}, and \texttt{\small GameActivity} are the most popular source for obtaining PII data. The largest amount of PII data flows to the Intent-related and Log-related sinks, such as \texttt{\small content.Intent} and \texttt{\small util.Log}. The use of the sink method may cause data leaks. For example, in a virtual social app, 
there is a data flow from \texttt{\small Location} method to \texttt{\small register\-Receiver()} in \texttt{\small content.Intent}. The BroadcastReceiver registered with the \texttt{\small register\-Receiver()} method is global and exportable by default. If access is not restricted, it can be accessed by any external app, passing Intent to it to perform specific functions. Therefore, dynamically registered BroadcastReceiver may lead to security risks such as denial of service attacks, APP data leakage, or unauthorized calls~\cite{Zhou2021}.

\begin{center}
\fcolorbox{black}{gray!10}{\parbox{\linewidth}{\textbf{Answer to RQ4}: A number of VR apps have the leakage risks of PII sensitive data. Most data flow from activity-related methods to intent-related methods.}}
\end{center}

\subsection{RQ5: How do the VR app developers comply with the privacy policies?}

As mentioned in \S~\ref{sec:privacy}, we collect policy statements from apps' privacy policies based on predefined ontologies by PolicyLint~\cite{Andow2019}. We check whether there are contradictory statements and GPDR violations in privacy policies and whether they are consistent with the app analysis results. 

\begin{figure}[h]
\centering
\includegraphics[width=0.45\textwidth]{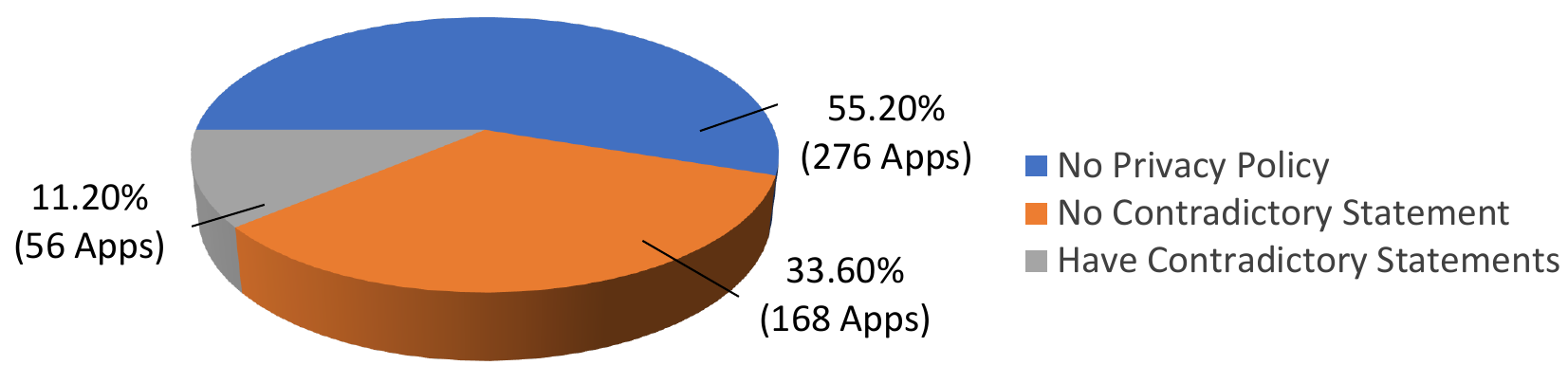}
\caption{App Privacy Policy Distribution}
\label{policy distribution}
\vspace{-4mm}
\end{figure}

As shown in Figure~\ref{policy distribution}, unexpectedly 276 apps (55.20\%) have no privacy policy though only 224 apps (44.80\%) have privacy policies. Among 224 apps with privacy policies, 56 of them contain contradictory statements. For example, in a business and finance VR app, the privacy policy states that they do not sell personal information to a third party while indicating that a third party may collect some specified category of personal information. It implies that there is still a certain percentage of VR apps containing unregulated privacy policies. Different from traditional mobile apps (Android apps on mobile phones), VR apps have higher chances to access highly-sensitive personal biometrics. Therefore, it is crucial to establish a consistent privacy policy for regulating VR app development.

\begin{table}[t]
\caption{GDPR Compliance Check}
\resizebox{0.95\linewidth}{!}{
\begin{tabular}{llc}
\toprule
\textbf{GDPR Violation Term} & \textbf{Risk Level} & \textbf{\# Apps} \\
\midrule
\rowcolor{mygray}
Missing Social Media Clause & High & 65 \\
\rowcolor{mygray}
Missing Data Sharing Information & High & 104  \\
\rowcolor{mygray}
Missing GDPR Roles & High & 133 \\
\rowcolor{mygray}
Missing Data Subject Rights & High & 60 \\
\rowcolor{mygray}
Missing Legal Ground & High & 134 \\
\rowcolor{mygray}
Missing Data Retention Information & High & 107 \\
\rowcolor{mygray}
Missing Timestamp & High & 45 \\
\rowcolor{mygray}
Missing National Authority & High & 109 \\
\rowcolor{mygray}
Missing Sections & High & 68 \\
\rowcolor{mygray}
Missing GDPR Specificity & High & 69 \\
Missing Dedicated Privacy Mailbox & Medium & 46 \\
Missing Data Security Information & Medium & 58 \\
Missing Top Level Link & Medium & 85 \\
\rowcolor{magicmint}
No Violation & N/A & 38 \\
\bottomrule
\end{tabular}
\label{GDPR Compliance}
}
\end{table}

Table~\ref{GDPR Compliance} reports the GDPR compliance check results. We find that only 38 apps among these 224 apps with privacy policies have privacy policies fully complying with GDPR. Meanwhile, there are 13 GDPR violation terms in our check results: 10 high-risk terms and 3 medium-risk terms. Most app privacy policies (i.e., 134) miss the legal ground. The results imply that the normality of privacy policies of VR apps still needs to be improved. Privacy policies with no compliance with the law may expose developers to legal risks.

\begin{figure*}[t]
\centering
\subfigure[Permission \& Biometrics Inconsistency Check]{
\begin{minipage}{0.45\textwidth}
\centering
\includegraphics[width=\textwidth]{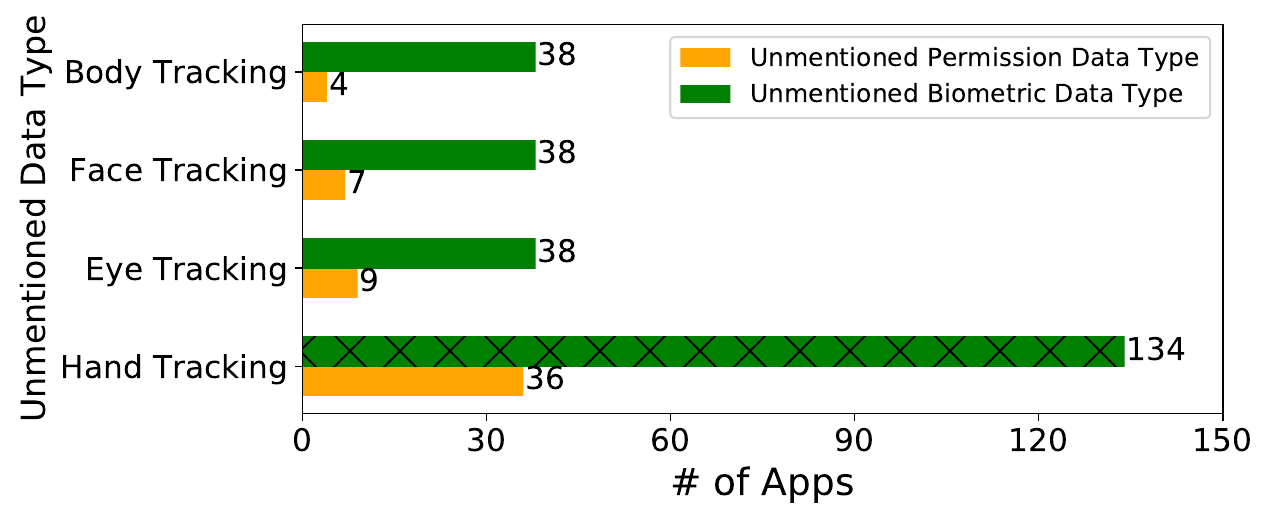}
\end{minipage}}
\subfigure[PII Inconsistency Check]{
\begin{minipage}{0.45\textwidth}
\centering
\includegraphics[width=\textwidth]{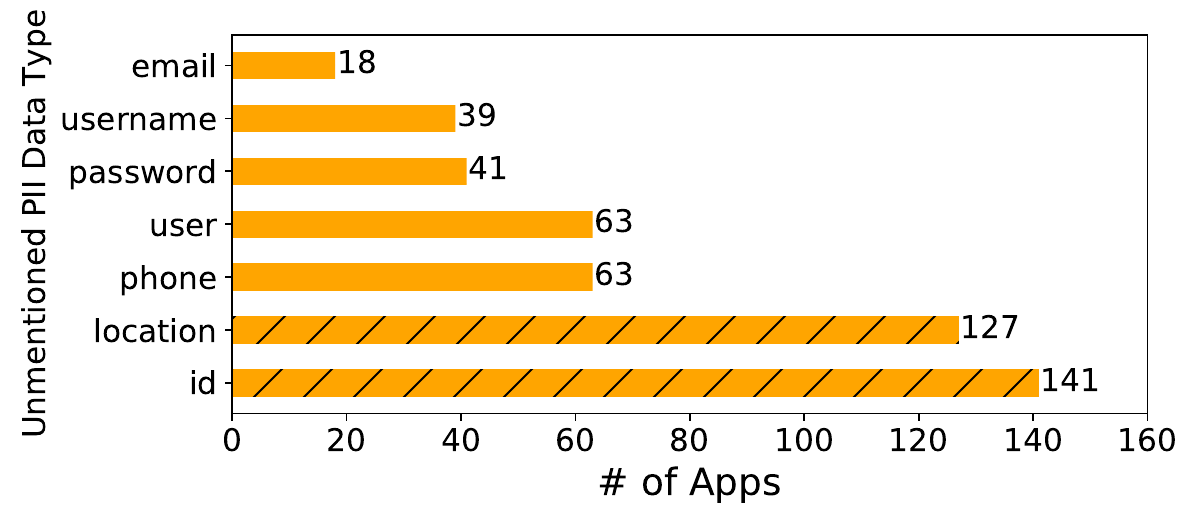}
\end{minipage}}

\caption{Privacy Policy Inconsistency Check}
\label{Inconsistency Check}
\vspace{-0.5cm}
\end{figure*}

In contrast to conventional mobile apps, VR apps need access to massive PII data, including not only device id, name, and phone number, but also additional highly-sensitive human biometrics, such as hand coordinates, eye rotation, body shape, and face expressions (\S~\ref{sec:privacy}). We further check whether accessing to this PII sensitive data is explicitly mentioned in the corresponding privacy policy. 
In particular, according to the result of the permission inconsistency check from the manifest analysis shown in Figure~\ref{Inconsistency Check}(a), we find that 41 apps with privacy policies do not mention the usage of hand, eye, body, and face data while they are found to request these permissions by the manifest analysis. Meanwhile, 36 of these apps use hand-tracking data but do not state so in their privacy policies. According to the result of the PII inconsistency check from the decompiled Java and Smali codes, shown in Figure~\ref{Inconsistency Check}(b), we find that most apps with privacy policies (183 apps) do not mention the use of PII data in detail though they are found to use PII collection method in the code analysis. Moreover, 141 of these apps used \textit{id} (e.g., device id) but do not mention it in their privacy policies. Further, we find that 136 apps have this inconsistency between biometric function usage and the privacy policy as shown in Figure~\ref{Inconsistency Check}(a). In addition, 134 of these apps identified hand-tracking data collection methods from the decompiled C\# codes but do not mention them in their privacy policies. The results show that there exist a number of irregularities in the privacy policies, which have not been updated in time to address the privacy concerns of VR app data collection.

\begin{center}
\fcolorbox{black}{gray!10}{\parbox{\linewidth}{\textbf{Answer to RQ5}: Less than half of VR apps offer privacy policies though 25.00\% of them contain contradictory statements. 
Meanwhile, 16.96\% of 224 VR apps with privacy policies comply with GDPR regulations, 81.70\% of them have no explicit mention of PII data usage in detail, and about 60\% have inconsistency between human biometrics collection and privacy policy.
}}
\end{center}

\subsection{Discussion}
We present two app cases to elaborate on security and privacy issues and offer several pieces of advice in VR app development.

\textbf{Case Study.} 
Considering the ethics issue, we have anonymized the specific names of the apps.
\textit{Case 1} (MD5 prefix: \texttt{\small 4dbd4\textasciitilde}) is a virtual social VR app with over 5,200 downloads on SideQuest . 
We find that this App has no root detection implementation. 
The lack of root detection may lead to SDI vulnerability and data breaches. We also find that this app includes the usage of SQL raw query function. 
Due to the lack of root detection, the attacker can directly use administrator privileges to access/modify the database, resulting in user data leakage. Moreover, it also adopts insecure random generators, 
which may cause a predictable random number. Attackers can use predictable values to bypass permission verification. This app also allows clear text traffic, incurring a risk that a cyber attacker could eavesdrop on the transmitted data. In addition, despite the identified usage biometric data collection API, 
we do not find the corresponding \texttt{\small <uses-permission>} tag 
in the \texttt{\small Android\-manifest.xml} file, implying that the app does not apply for the permission while using the corresponding APIs. 
We also do not find a statement in our privacy policy with respect to the collection of such biometric data. The \texttt{id}, \texttt{phone}, and \texttt{location} data collection are not mentioned in the privacy policy while are adopted in the source code. In addition, the privacy policy contains GDPR violations, such as missing data sharing information and missing top-level link.

\textit{Case 2} (MD5 prefix: \texttt{\small c2300\textasciitilde}) is another virtual social VR app with more than 9,500 downloads . 
We find that the \texttt{\small allow\_backup} flag in the manifest file is marked as \textit{True}, 
which leads to a data leak risk. We also find that there exist some insecure encrypt functions by matching pattern \texttt{\small AES/ECB}, which indicates the app uses insecure ECB mode in the Cryptographic encryption algorithm. 
Besides, this app also utilizes some insecure hash functions including MD5-related hash functions (e.g., \texttt{\small Encrypter.MD5}
), SHA-1-related functions (e.g., \texttt{\small Util.sha1}) and so on. A tracker called \textit{Unity3d Ads} is also identified in this app. Besides, it lacks root detection. From the taint analysis result, we find that there exists PII data flow from \texttt{\small PackageManager} function to \texttt{\small Log} function. The use of \texttt{\small Log} function may expose network packet data to attackers and thus be intercepted by attackers for illegal activities in the metaverse such as harassment. The \texttt{location}, \texttt{password}, \texttt{user}, \texttt{username}, \texttt{phone}, and \texttt{location} data collection are not mentioned in the privacy policy but they are adopted in the source code.

\textbf{Development Advice}. Considering the findings in the case study, we offer some advice on the development of VR apps.

\noindent
\colorbox{gray!30}{\it Advice 1: Set proper secure flags in the manifest file.}
Before releasing the app, it is suggested to set the security-related flags in the \texttt{\small Androidmanifest.xml} file as \textit{False}, such as \texttt{\small allow\_backup}, \texttt{\small debuggable}, and  \texttt{\small use\_cleartext\_traffic}. If these flags are \textit{True}, users' private behaviors can be inspected by attackers.

\noindent
\colorbox{gray!30}{\it Advice 2: Enable root detection when starting the VR app.} 
A rooted device may cause the association of app data and user data~\cite{Sun2015}. Consequently, invaded malware lurking in the VR device can steal users' private information. This can be even more dangerous in the metaverse, which involves users' frequent interactions (e.g., transactions of virtual assets).

\noindent
\colorbox{gray!30}{\it Advice 3: Do not use insecure hash functions/encryption algorithms.}
Compared with traditional mobile apps, VR apps collect more diverse data (e.g., video and voice). It is crucial to use effective encryption methods, such as secure hash functions (e.g., SHA-256 \cite{Appel2015}) and encryption methods, such as RSA with OAEP padding.

\noindent
\colorbox{gray!30}{\it Advice 4: Check data flows used by the trackers.}
It is difficult to guarantee the security of these user data sent to third-party platforms by trackers. Therefore, developers should check the data flows used by the trackers to ensure no abuse or misuse.

\noindent
\colorbox{gray!30}{\it Advice 5: Comply with permission requests to collect biometric data.}
Developers should comply with specifications for sensitive data collection, while app stores should strengthen code audits to prevent similar malware releases. 
Meanwhile, the sensitive data collection functions should be regulated to ensure the user's right to know how sensitive data is collected and prevent it from being misused.

\noindent
\colorbox{gray!30}{\it Advice 6: Adapt privacy policies to fulfill VR apps' new features.} 
Developers need to develop new privacy policies according to VR apps' new features, such as privacy concerns with immersive social interactions and the collection of human biometrics. Further, the publication of a privacy policy is subject to relevant legislation.

\vspace{-0.3cm}
\section{Limitation and Threats to Validity}

\textbf{Limitation.}
The limitation of our research lies in the integrity of decompiled code. Since some apps are shelled for anti-cheating purposes, we cannot get the complete decompiled code to analyze the API call relationships in all apps. In addition, the data transmissions between different data-collection functions and the tracking algorithms is not open-sourced in the Oculus developer documentation. We will further analyze the call chains of these API functions in the future, especially for those of biometric data collection.

\textbf{Threats to External Validity.}
The threats to external validity limit the scalability of our approach. In our biometric data propagation analysis, we mainly focus on VR apps developed based on Unity though there are other frameworks, such as UE~\cite{Qiu2016}, libGDX \cite{Stemkoski2015}, and so on. Moreover, we only analyze VR apps working on Meta Oculus Quest 2 while there are some other popular VR/AR devices, such as the HTC VIVE Pro 2, Sony Playstation VR 2, and Pico 4, many of which are also Android or its variants (our tool may also apply to). 
In short, more types of VR apps need to undergo security and privacy assessments in the future.

\textbf{Threats to Internal Validity.}
Although we collect 500 VR apps, which are almost 3 $\times$ of the state-of-the-art tool \textsc{OVRseen}~\cite{Trimananda2022}, we will evaluate more VR apps with the proliferation of VR and metaverse. 
Moreover, as for the static analysis for OS-related security and privacy vulnerabilities, we referred to pre-defined patterns based on \cite{Sun2021} and the precision result they claimed is 96.19\%. We manually checked the result of 20\% of the apps and found no false positives, thus reducing the impact of tool accuracy. The taint analysis adopted in this work can also lead to some false positives, thereby affecting the accuracy of the results. To address this issue, we manually checked 20\% of the identified paths and found no false positives, thus mitigating its side effect. With respect to another internal validity threat caused by the accuracy of the GDPRWise, we manually sampled 50 privacy policy cases to verify the accuracy of their GDPR compliance detection and found 9 false positives. We hypothesize three possible reasons: (i) the privacy policy is written in a non-English language (e.g., Japanese), which affects its detection accuracy; (ii) the link to the privacy policy is in PDF format, which is not supported by the service; and (iii) the link to the privacy policy contains additional external links. The extent of the impact needs further investigation.

\vspace{-0.2cm}
\section{Related Work}

\textbf{Program Analysis of Mobile Apps.}
Many recent studies adopt both static tools and dynamic techniques to analyze the security and privacy vulnerabilities of mobile apps. Lee et al. \cite{Lee2016} proposed a static tool to analyze inter-communication between Android Java and JavaScript codes.  
Sandeep \cite{Sandeep2019} combined deep learning and static analysis to detect Android malware with high accuracy. Regarding dynamic analysis, Reardon et al. \cite{Reardon2019} constructed a testing environment to detect whether the app bypassed the permission model to access protected data.  
Huang et al. \cite{Huang2019} proposed a testing framework based on net packet fuzzing for Android apps. 
In this paper, we use static analysis combined with privacy-policy analysis to perform security and privacy analysis on emerging VR apps.

\textbf{Security and Privacy Analysis of VR Apps.}
With the rapid development of VR devices and metaverse platforms, the analysis of VR apps has received increasing attention. For example, Trimananda et al. \cite{Trimananda2022} proposed a method, namely \textsc{OVRseen} to analyze the privacy policies in Oculus VR apps by collecting network traffic and comparing them with the privacy policies although its dynamic analysis also has limited coverage for execution paths and unsoundness as indicated in~\cite{David2018}. Yarramreddy et al. \cite{Yarramreddy2018} proposed a forensic analysis of VR social apps to reveal some forensically relevant data from network traffic and the VR systems. Casey et al. \cite{Casey2021} discovered a new attack against VR systems, which can open the VR camera without user permission and insert images into users' vision to distract users' attention in a virtual environment. 
This paper focuses on metaverse-related VR apps with a comprehensive assessment of their security and privacy status.

\textbf{Unity-based VR Apps.}
Since most VR apps have been developed based on Unity to render 3D environment, achieve immersive user experience, and collect biometric data, we also review related work as follows. Shim et al. \cite{Shim2018} proposed a reverse engineering method with a combination of static and dynamic analysis to analyze malicious Unity apps. This tool can be used to analyze the native code of Java, C, C++, and the Mono layer where the C\# code runs. Volokh et al. \cite{Volokh2022} proposed a Unity game code logic analysis tool based on static analysis, which can be used to provide an available action state set at a game state for players. Zuo et al. \cite{Zuo2022} conducted an in-depth analysis of the security of paid implementations in Unity-based handheld games by designing and implementing the static tool, namely PaymentScope to automatically identify vulnerable IAPs in mobile games.
In this paper, we design a variant of PaymentScope to detect not only vulnerable IAPs but also biometric data usage.

\vspace{-0.3cm}
\section{Conclusion}
With the proliferation of diverse VR devices and the increasing attention of the metaverse in recent years, VR apps have received a boosted development and proliferation. Although numerous VR apps have been released, little attention has been paid to the security and privacy issues of emerging VR apps. In this paper, we have developed a security and privacy assessment tool, namely the VR-SP detector for VR apps. The VR-SP detector has been implemented with the integration of program static analysis and privacy policy analysis methods. Using the VR-SP detector, we have conducted the security and privacy assessment of 500 popular VR apps. Our analytical results have revealed important security and privacy issues of existing metaverse-related VR apps. Based on our findings, we have made development recommendations for future VR apps with security and privacy preservation.

\vspace{-0.3cm}
\begin{acks}
The work described in this paper is partially supported by the National Natural Science Foundation of China (62032025), COMP Department Start-up Fund of Hong Kong Baptist University (HKBU), Faculty Start-up Grant for New Academics of HKBU, SD/COMP Joint Research Scheme (ID: P0042739), and Departmental Incentive Scheme of HKBU COMP. We would like anonymous reviewers for their constructive comments.
\end{acks}

\bibliographystyle{ACM-Reference-Format}
\bibliography{reference}
\end{document}